\documentclass[
prd,
reprint,
showpacs,
superscriptaddress,
amssymb,
floatfix
]{revtex4-1}

\usepackage{latexsym}
\usepackage{amsfonts, amssymb, amsmath}
\usepackage{graphicx}
\usepackage{epstopdf}
\usepackage{bm}% bold math
\usepackage{color}
\usepackage{CJK}
\newcommand{\SMOMg}{\text{RI/SMOM}_{\gamma_\mu}}
\newcommand{\MSbar}{\overline{\text{MS}}}
\DeclareMathOperator{\Tr}{Tr} 
\newcommand{\Xsl}[1]{\raise.15ex\hbox{/}\kern-.57em #1}

\begin{document}

\begin{CJK*}{UTF8}{}

\title{Nucleon mass and isovector couplings in 2+1-flavor dynamical domain-wall lattice QCD near physical mass}
\date{\today}

%\author{author list}
\author{Michael Abramczyk}
\affiliation{Physics Department, University of Connecticut, Storrs, CT 06269-3046}
\author{Thomas Blum}
\affiliation{Physics Department, University of Connecticut, Storrs, CT 06269-3046}
\affiliation{RIKEN-BNL Research Center, Brookhaven National Laboratory, Upton, NY 11973}
\CJKfamily{min}
\author{Taku Izubuchi (出渕卓)}
\affiliation{Physics Department, Brookhaven National Laboratory, Upton, NY 11973}
\affiliation{RIKEN-BNL Research Center, Brookhaven National Laboratory, Upton, NY 11973}
\author{Chulwoo Jung}
\affiliation{Physics Department, Brookhaven National Laboratory, Upton, NY 11973}
\affiliation{RIKEN-BNL Research Center, Brookhaven National Laboratory, Upton, NY 11973}
\author{Meifeng Lin}
\affiliation{Computational Science Initiative, Brookhaven National Laboratory, Upton, NY 11973}
\author{Andrew Lytle}
\affiliation{INFN, Sezione di Roma Tor Vergata, Via della Ricerca Scientifica 1, 00133 Roma RM, Italy}
\CJKfamily{min}
\author{Shigemi Ohta (太田滋生)}
\affiliation{Theory Center, The High Energy Accelerator Research Organization (KEK), Oho 1-1, Tsukuba, Ibaraki 3050801, Japan}
\affiliation{Deaprtment of Particle and Nuclear Physics, The Graduate University of Advanced Studies (SOKENDAI), Hayama, Kanagawa 2400193, Japan}
\affiliation{RIKEN-BNL Research Center, Brookhaven National Laboratory, Upton, NY 11973}
\email{shigemi.ohta@kek.jp}
\CJKfamily{min}
\author{Eigo Shintani (新谷栄悟)}
\affiliation{RIKEN Advanced Institute of Computational Sciences (AICS), Kobe, Hyogo 650-0047, Japan}
\collaboration{for RBC and UKQCD collaborations}
\pacs{12.38.Gc,14.20.Dh,11.15.Ha}

\begin{abstract}
%\vspace{-77mm}\parbox{0.85\textwidth}{\flushright\rm \hfill KEK-TH-2167, RBRC-1320}\vspace{70mm}
We report nucleon mass, isovector vector and axial-vector charges, and tensor and scalar couplings,  calculated using two recent 2+1-flavor dynamical domain-wall fermions lattice-QCD ensembles generated jointly by the RIKEN-BNL-Columbia and UKQCD collaborations.
These ensembles were generated with Iwasaki \(\times\) dislocation-suppressing-determinant-ratio gauge action at inverse lattice spacing of 1.378(7) GeV and pion mass values of 249.4(3) and 172.3(3) MeV.
The nucleon mass extrapolates to a value \(m_N = 0.950(5)\) GeV at physical point.
The isovector vector charge renormalizes to unity in the chiral limit, narrowly constraining excited-state contamination in the calculation.
The ratio of the isovector axial-vector to vector charges shows a deficit of about ten percent.
The tensor coupling no longer depends on mass and extrapolates to 1.04(5) in \(\overline {\rm MS}\) 2-GeV renormalization at physical point, in a good agreement with the value obtained at the lightest mass in our previous calculations and other calculations that followed.
The scalar charge, though noisier, does not show mass dependence and is in agreement with other calculations.
\end{abstract}

\maketitle
\end{CJK*}

\section{Introduction} 

The RIKEN-BNL-Columbia (RBC) collaboration and subsequently joint RBC and UKQCD collaborations have been investigating nucleon structure using the domain-wall fermions (DWF) quarks on a sequence of quenched \cite{Sasaki:2003jh,Orginos:2005uy}, 2- \cite{Lin:2008uz}, and 2+1-flavor \cite{Yamazaki:2008py,Yamazaki:2009zq,Aoki:2010xg} dynamical DWF ensembles at various mass values \cite{Blum:2000kn,Aoki:2004ht,Allton:2008pn,Aoki:2010dy,Arthur:2012yc,Blum:2014tka}.
As is well known, the DWF scheme allows us to maintain continuum-like flavor and chiral symmetries on the lattice, and it helps to simplify non-perturbative renormalizations \cite{Martinelli:1994ty,Blum:2001sr,Aoki:2007xm,Sturm:2009fk}.
In this paper we report nucleon isovector vector and axial-vector charges and tensor, \(g_T^{u-d}=\langle 1 \rangle_{\delta u - \delta d}\), and scalar, \(g_S^{u-d}\), couplings calculated using two recent 2+1-flavor dynamical DWF lattice-QCD ensembles generated jointly by the RBC and UKQCD collaborations with Iwasaki \(\times\) dislocation-suppressing-determinant-ratio (DSDR) gauge action at \(\beta=1.75\), \(a^{-1}=1.378(7)\) GeV, and pion mass of 249.4(3) and 172.3(3) MeV \cite{Arthur:2012yc,Blum:2014tka}, which are significantly lighter than our previous nucleon works.

In our earlier works calculated with degenerate up- and down-quark mass set at considerably heavier than physical values \cite{Yamazaki:2008py,Yamazaki:2009zq,Aoki:2010xg}, we observed the vector-current form factors behaving reasonably well in trending toward experiments: both Dirac and Pauli mean-squared charge radii and the isovector anomalous magnetic moment appeared to linearly depend on the pion mass squared.
The radii extrapolated to the physical mass undershoot the experimental value by about 25 percent \cite{Pohl:2010zza,Bezginov:2019mdi}.
It would have been interesting to see if the present calculation confirmed this earlier trend, or if it could help resolve the discrepancy seen between muon Lamb shift experiment \cite{Pohl:2010zza} and older electron scattering ones.
However, our current numerical precision from relatively small statistics and large momentum transfer is yet to be competitive with Lamb shift experiments \cite{Pohl:2010zza,Bezginov:2019mdi} which now are prevailing \cite{PhysRevD.98.030001,Xiong:2019}.
So we would like to defer reporting our form factors at finite momentum transfers until a future date when we will have better precision. 

In our earlier calculation of axial-vector current form factors, we saw a significant deficit in the calculated axial-vector charge, \(g_A\), and form factors in general appear more susceptible to finite-size effects than the vector-current ones \cite{Yamazaki:2008py,Yamazaki:2009zq}.
We find in our present calculations at lighter pion mass that this deficit persists, and we investigate potential causes for this in detail in Section V.

In contrast, the isovector tensor coupling showed interesting downward departure at the lightest mass to about 1.0 away from the flat higher-mass values of about 1.1 \cite{Aoki:2010xg}.
Whether this trending continues in our present calculations at considerably lighter mass is obviously an interesting question.
The tensor and scalar couplings are also relevant to the search for new physics beyond the standard model such as neutron electric dipole moment \cite{Bhattacharya:2016zcn,Yoon:2016jzj}.

Some preliminary analyses of these nucleon observables have been reported at recent Lattice conferences \cite{Lin:2014saa,Ohta:2013qda,Ohta:2014rfa,Ohta:2015aos,Abramczyk:2016ziv}.
In addition, the LHP collaboration also calculated some nucleon observables \cite{Syritsyn:2009mx} using a RBC+UKQCD 2+1-flavor dynamical DWF ensemble \cite{Allton:2008pn}.

\section{Numerical Method} 

In this paper we concentrate on our results for mass and four isovector couplings of nucleon: vector and axial-vector charges and tensor and scalar couplings.
Though we summarize their definitions and computational methods later in this section, we refer the readers to our earlier publication \cite{Yamazaki:2009zq} for the full details.

The two 2+1-flavor dynamical domain wall fermion gauge field ensembles we use in this work \cite{Arthur:2012yc,Blum:2014tka} were generated jointly by the RBC and UKQCD collaborations on \(32^3\times64\) four-dimensional volume and \(L_s=32\) in the fifth dimension with Iwasaki \(\times\) DSDR gauge action at the gauge coupling of \(\beta=1.75\) and 2+1-flavors of dynamical DWF quarks of the domain-wall height of 1.8, strange-quark mass of 0.045, and degenerate up- and down-quark mass of 0.0042 and 0.001 in lattice units.
These parameters result in the inverse lattice spacing, \(a^{-1}\), of 1.378(7) GeV  and the DWF residual quark mass of 0.001842(7).
Note that the inverse lattice spacing has been slightly revised from the original \cite{Aoki:2010dy,Arthur:2012yc} using the global chiral and continuum fits in conjunction with new physical-mass ensemble sets \cite{Blum:2014tka} with M\"obius DWF quarks.
Thus the heavier of the two ensembles corresponds to the pion mass, \(m_\pi\), of 249.4(3) MeV and spatial lattice extent \(L\) of \(m_\pi L = 5.79(6)\), and the lighter to 172.3(3) MeV and 4.00(6), respectively.
Our measurement calculations were made using 165 configurations between the molecular-dynamics (MD) trajectory 608 and 1920 with an 8-trajectory interval for the former, and using 39 configurations between 748 and 1420 with a 16-trajectory interval for the latter.

We refer to our earlier publications  \cite{Lin:2008uz,Yamazaki:2008py} for the details of two- and three- point correlation functions for the nucleon.
A conventional nucleon operator, \( N(x) = \epsilon_{abc} \left\{[u_a(x)]^T C \gamma_5 d_b(x)\right\}u_c(x),\) with color indices \(a\), \(b\), and \(c\), quark flavors \(u\) for up and \(d\) for down, and charge conjugation operator \(C=\gamma_4 \gamma_2\) is used.
Additionally, gauge-invariant Gaussian smearing \cite{Alexandrou:1992ti,Berruto:2005hg} is applied to suppress excited-state contamination:
For both ensembles we compared the Gaussian width of 4 and 6 lattice units and found
the wider, width-6, results settle on a plateau more quickly, and the narrower width-4 results merge with them.
From these observations we concluded the wider width-6 Gaussian smearing is sufficient for our study.

Isospin symmetry is enforced for the up and down quarks, and we calculate only proton matrix elements of the third isospin component of the vector and axial-vector currents,
\begin{eqnarray}
\langle P | V_\mu^3(x) |P\rangle &=&
\langle P |
\bar{u}(x) \gamma_\mu u(x) -\bar{d}(x) \gamma_\mu d(x) | P\rangle,
\\
\langle P | A_\mu^3(x) |P\rangle &=&
\langle P |
\bar{u}(x) \gamma_5 \gamma_\mu u(x) -\bar{d}(x) \gamma_5 \gamma_\mu d(x) | P\rangle.\nonumber\\
\end{eqnarray}
All quark-disconnected diagrams cancel in these matrix elements.

We use a source-sink separation of 9 lattice units, or about 1.3 fm.
This is sufficiently long for the observables reported in this paper to be free of excited-state contamination, as is demonstrated below with augmentative calculations with source-sink separation of 7 lattice units \cite{Ohta:2013qda,Ohta:2014rfa}.
For the two point correlation functions we use the same Gaussian-smeared sources and point or Gaussian-smeared sinks:
We will refer to the former as G-L and the latter as G-G respectively.
We also calculate the tensor coupling, \( g_T=\langle 1 \rangle_{\delta q}\), and scalar coupling, \(g_S\), the same way.
We used a conventional measurement strategy for the former with seven source-sink pairs for each configuration, and an ``all-mode-averaging (AMA)'' strategy \cite{Blum:2012my} for the latter with 112 sloppy solves with sources at two sets of 8 evenly spaced spatial locations, from (0,0,0) to (16, 16, 16) and from (8,8,8) to (24,24,24)  for  \(T=0\), 8, 16, 24, 32, 40, and 48 and four precise solves at spatial origins with \(T=0\), 16, 32 and 48 for each configuration.
In addition, to test for excited state contamination, calculations with source-sink separation of 7 lattice units were made with 64 sloppy solves with sources at 8 evenly spaced spatial locations  with \(T=0\), 8, 16, 24, 32, 40, 48, and 56 and  and one precise solve at spatial originswith \(T=0\).

\section{Nucleon mass}

\begin{figure}[tb]
\includegraphics[width=\columnwidth,clip]{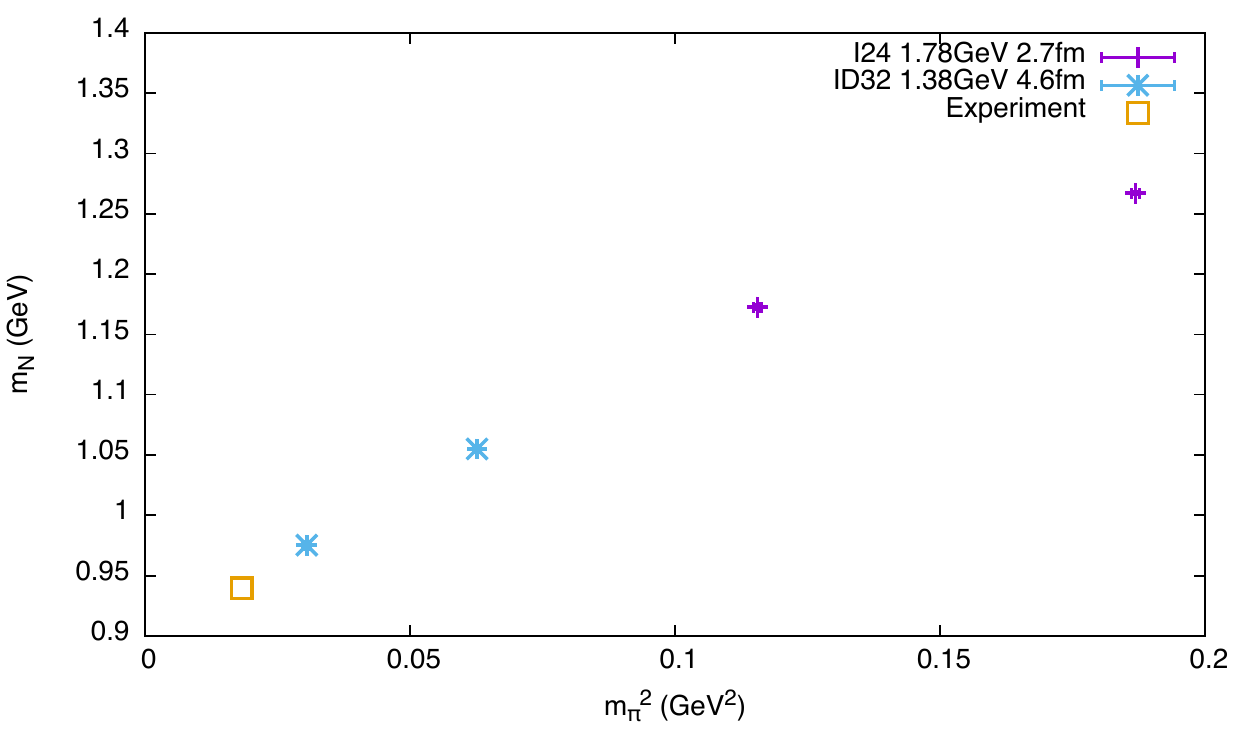}
\caption{
\label{fig:mpi2mN}
Estimated nucleon mass, \(m_N\), plotted against estimated pion mass squared, \(m_\pi^2\), of the present ensembles (ID32, cyan) and the two lightest of ref.\ \cite{Yamazaki:2009zq,Aoki:2010xg} (I24, magenta) with the new and more accurate estimate for the inverse lattice spacing \cite{Blum:2014tka}.
The present results linearly extrapolate to the experiment (\(\Box\)) within the statistical error.
}
\end{figure}

Table~\ref{tab:energies}
\begin{table}[b]
\caption{
\label{tab:energies}
Fitted nucleon energy obtained from the Gaussian-smeared sink two-point correlation function (G-G).
}
\begin{tabular}{lllll}
\hline
\(mm_qa\) & \(|\vec p|^{2}\) &  fit range & energy & \(\chi^{2}\)/dof \\
\hline\hline
0.001 & 0 & 5-10 & 0.707(5) & 0.33 \\
0.001 & 1 & 5-10 & 0.734(5) & 0.13 \\
0.001 & 2 & 5-10 & 0.761(6) & 0.16 \\
0.001 & 3 & 5-10 & 0.786(6) & 0.10 \\
0.001 & 4 & 5-10 & 0.807(8) & 0.72 \\
0.0042 & 0 & 6-12 & 0.765(12) & 1.55 \\
0.0042 & 1 & 6-12 & 0.790(13) & 1.62 \\
0.0042 & 2 & 6-12 & 0.816(14) & 2.01 \\
0.0042 & 3 & 6-12 & 0.84(2) & 2.20 \\
0.0042 & 4 & 6-12 & 0.86(2) & 1.77 \\
\hline
\end{tabular}
\end{table}
summarizes the nucleon energies obtained from correlated, single-exponential fits to the G-G two point correlation function.
To improve statistics the correlation function is averaged over forward and backward propagating nucleon and anti-nucleon states.
The fit ranges were chosen after inspecting the effective masses and taking the shortest distance from the source consistent with an acceptable \(\chi^2\) value from the fit.
Increasing this minimum distance by a single time unit does not change the energy beyond the statistical error.
From these we estimate the nucleon mass for the present ensembles as summarized in Tab.\ \ref{tab:mN}
\begin{table}[h]
\caption{
\label{tab:mN}
Nucleon mass estimates.  We also list the revised numbers for the two lightest ensembles of ref.\ \cite{Yamazaki:2009zq,Aoki:2010xg} with the new and more accurate estimate for the inverse lattice spacing \cite{Blum:2014tka}.
}
\begin{tabular}{llll}
\hline
\multicolumn{1}{c}{\(a^{-1}\)[GeV]}&
\multicolumn{1}{c}{\(m_qa\)} &
\multicolumn{1}{c}{\(m_Na\)} &
\multicolumn{1}{c}{\(m_N\) [GeV]}\\
\hline\hline
1.378(7) & 0.001 & 0.7077(8) & 0.9752(11)\\
               & 0.0042 & 0.7656(2) & 1.0550(3)\\
\hline
1.7848(5)  & 0.005 & 0.6570(9)& 1.1726(16)\\
                 & 0.01    & 0.7099(5) & 1.2670(9)\\
\hline
\end{tabular}
\end{table}
and Fig.\ \ref{fig:mpi2mN}.
Whereas our prior calculations at heavier pion masses found nucleon mass extrapolating to values much higher than experiment, with pion masses now sufficiently close to physical that the new data extrapolates, linearly in terms of pion mass squared, \(m_\pi^2\), to a value \(m_N = 0.6894(7) a^{-1} = 0.6894(7)\times1.378(7) {\rm GeV} = 0.950(5)\) GeV.
This extrapolation is only slightly more than two standard deviations away from the average of proton and neutron mass, 0.938918747(6) GeV \cite{PhysRevD.98.030001}.
The slope in this linear extrapolation is steeper than that observed in our earlier calculations with heavier mass:
the result constrains non-linear dependence of nucleon mass on pion mass squared.

\section{Vector Charge}

\begin{figure}[t]
\includegraphics[width=\columnwidth]{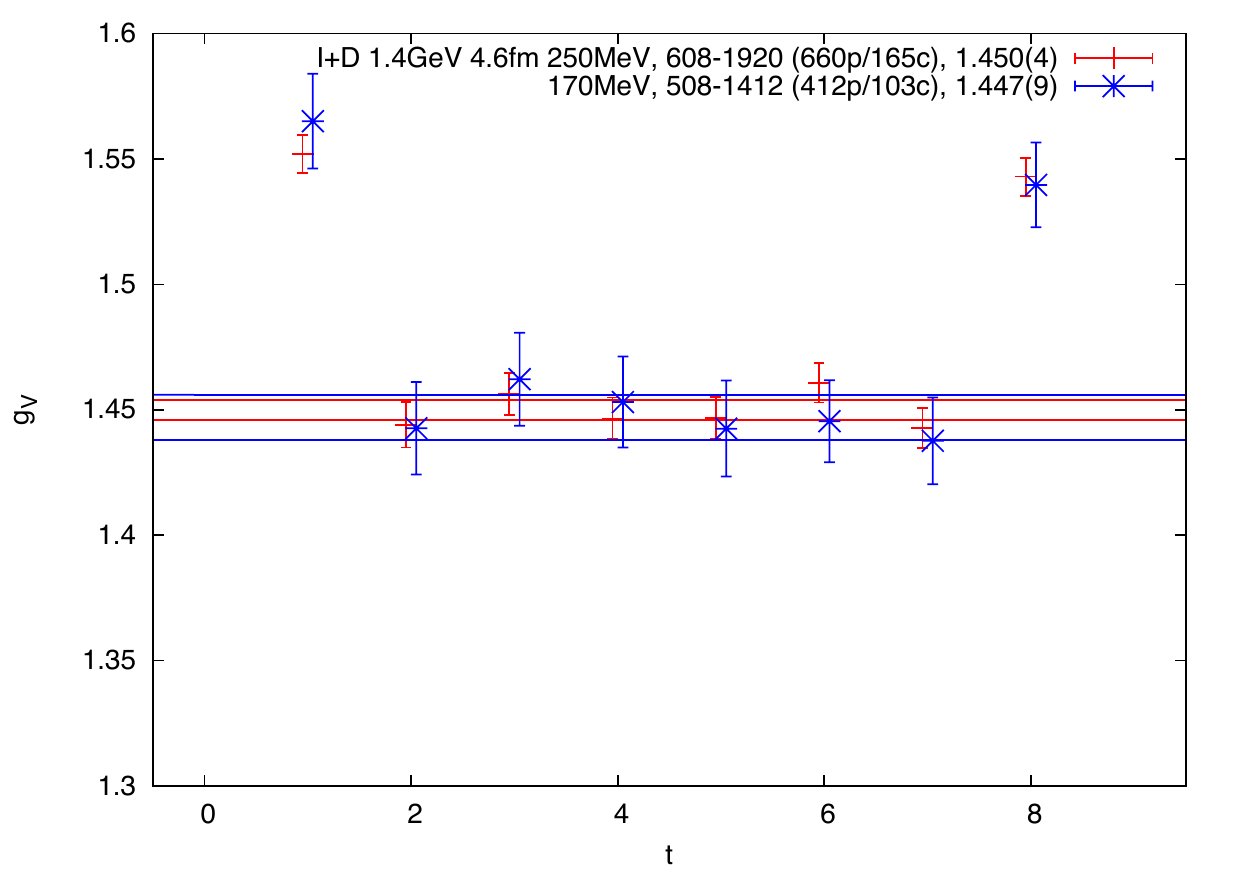}
\caption{
\label{fig:gVplateau}
Isovector vector charge, \(g_V\), from \(m_\pi=172.3(3)\) and 249.4(3) MeV ensembles.}
\end{figure}

Signals for the isovector vector charge, \(g_V\), are shown in Fig.\ref{fig:gVplateau}
for both \(m_\pi=172.3(3)\) and 249.4(3) MeV ensembles.
Robust time-independent plateaux are seen.
In the following we average values in the range \(3\le t_{\rm op}\le 6\),  and find 1.450(4) for the heavier and 1.447(9) for the lighter ensemble, respectively.
The values compare with the inverse of the vector-current renormalization, \(Z_V\), computed in the meson sector \cite{Arthur:2012yc}, \(0.664(5)^{-1} = 1.506(11)\) and \(0.669(4)^{-1} = 1.495(9)\).
Alternatively we linearly extrapolate the two calculated vector-charge values of 1.450(4) at \(m_fa=0.0042\) and 1.447(9) at 0.001 to the chiral limit, \(m_fa = -m_{\rm residual}a = -0.0018427\) to obtain a value 1.474(11).
This is to be compared with the inverse of the meson-sector vector-current renormalization, \(Z_V\),  in the chiral limit, \(Z_V^{-1} = 0.673(8)^{-1} = 1.49(2).\)
Thus the nucleon vector-charge in the chiral limit renormalizes to unity within statistical errors: \(Z_V g_V = 0.992(14).\)

\begin{figure}[b]
\begin{center}
\includegraphics[width=\columnwidth,clip]{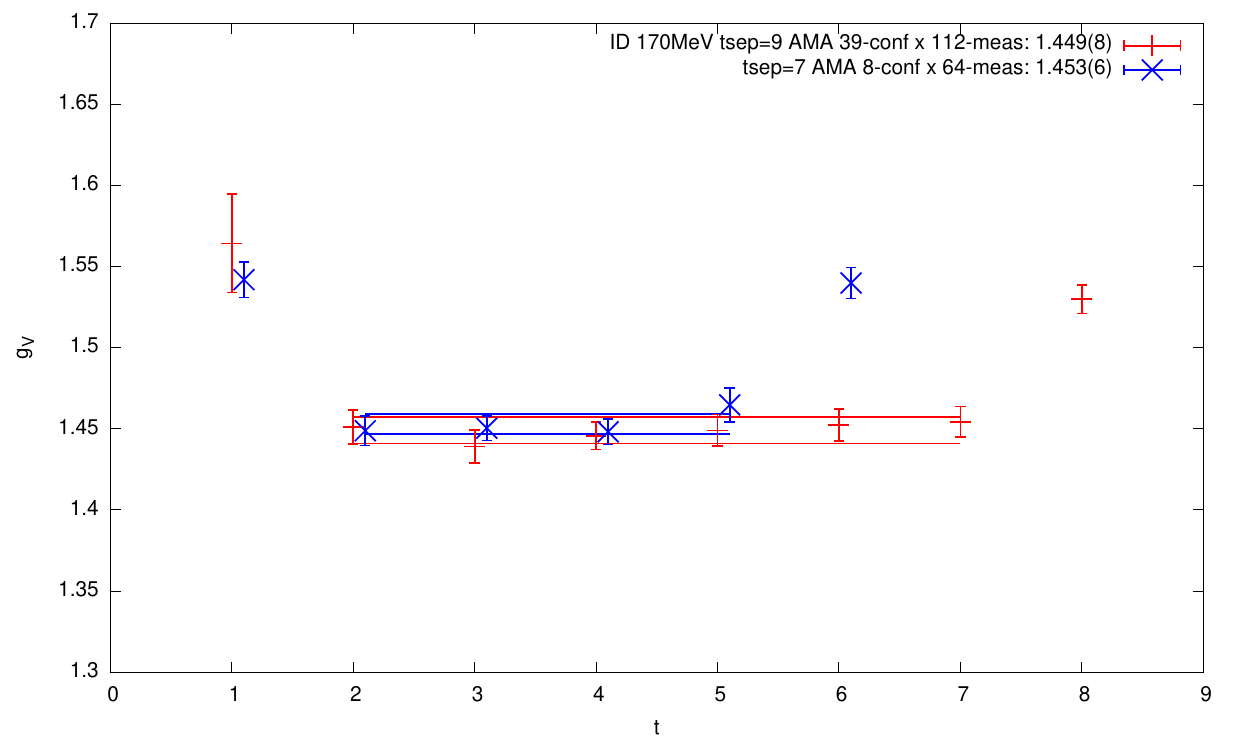}
\end{center}
\caption{
\label{fig:7+9}
Signals for isovector vector charge, \(g_V\), from source-sink separation of 7 and 9 lattice units agree to within 0.6 percent, well within statistical errors.
}
\end{figure}

\begin{figure}[t]
\begin{center}
\includegraphics[width=\columnwidth,clip]{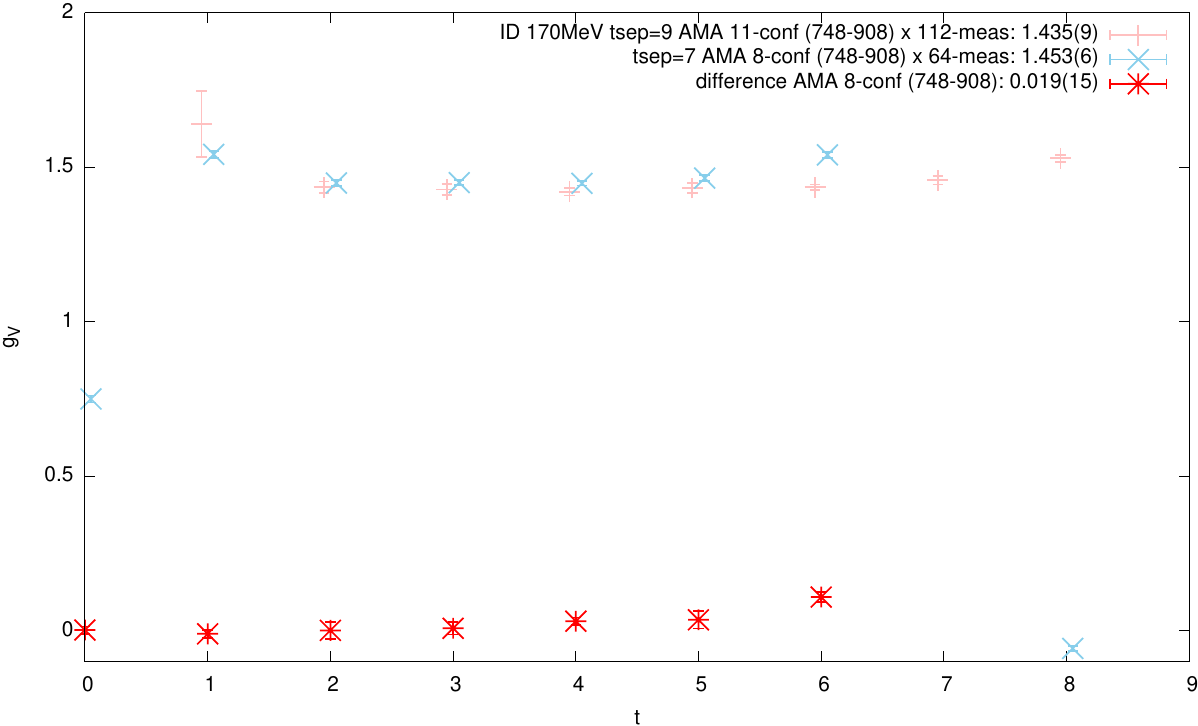}
\end{center}
\caption{
\label{fig:7-9}
Direct comparison of signals for isovector vector charge, \(g_V\), from source-sink separation of 7 and 9 lattice units agree for trajectories 748, 764, 796, 828, 844, 876, 892, and 908 for 7 and three more in the same range for 9.  The difference is consistent with zero except for separation 6 where a larger DWF fifth-dimensional effect can be seen for 7.
}
\end{figure}

Had the renormalized vector charge, \(Z_V g_V\), deviated from unity, a likely cause would be excited-state contamination, \(\sum_{\rm excited states} \langle {\rm excited state} | g_V | 0\rangle\), through violation of vector-current conservation at \(O(a^2)\) in lattice spacing \(a\).
That our vector charge renormalizes to unity within a couple of percent statistical error constrains such excited-state contamination.
As is shown in Fig.\ \ref{fig:7+9},
signals for isovector vector charge, \(g_V\), from source-sink separation of 7 and 9 lattice units agree to within 0.6 percent, well within statistical errors.
If we assume the first excited state is the ground-state nucleon plus a pion of mass \(m_\pi a = 0.1249(2)\) with a unit of lattice momentum, \(2\pi/L = 0.1963\), then it decays as \(\exp(-2\times(0.1249+0.1963)) = 0.5260\) in two lattice units of time from 7 to 9.
So the relative amplitude of this state in the source times the \(O(a^2)\) mixing matrix element cannot exceed approximately one percent.
It may appear the statistics used for this comparison is small, but it is sufficient in confirming  that the results from the short and long source-sink separations in agreement with each other.
This is because the short- and long separation results are correlated within each configuration.
Indeed such a correlation can be directly studied by limiting the comparison to configuration between trajectory 748 and 908:
there are all the eight configurations for separation 7, and additional three more for 9.
The jack-knife difference between the two separations are consistent with zero except for separation 6 where a larger DWF fifth-dimensional effect is expected in the separation 7 result (see Fig.\ \ref{fig:7-9}.)
Hence we conclude that there is no evidence for excited-state contamination detected in the most precise of our observables, the vector charge, in the lighter ensemble.
In addition the source smearing study done for both ensembles did not suggest any different behavior in the heavier and the lighter ensemble should be denser in excitation spectrum.
Accordingly possible excited-state contamination in the heavier ensemble must be smaller than in the lighter as the excitations are farther away.
There is no evidence for excited-state contamination in our ensembles.
\section{axial-vector charge}

\begin{figure}[tb]
\begin{center}
\includegraphics[width=\columnwidth]{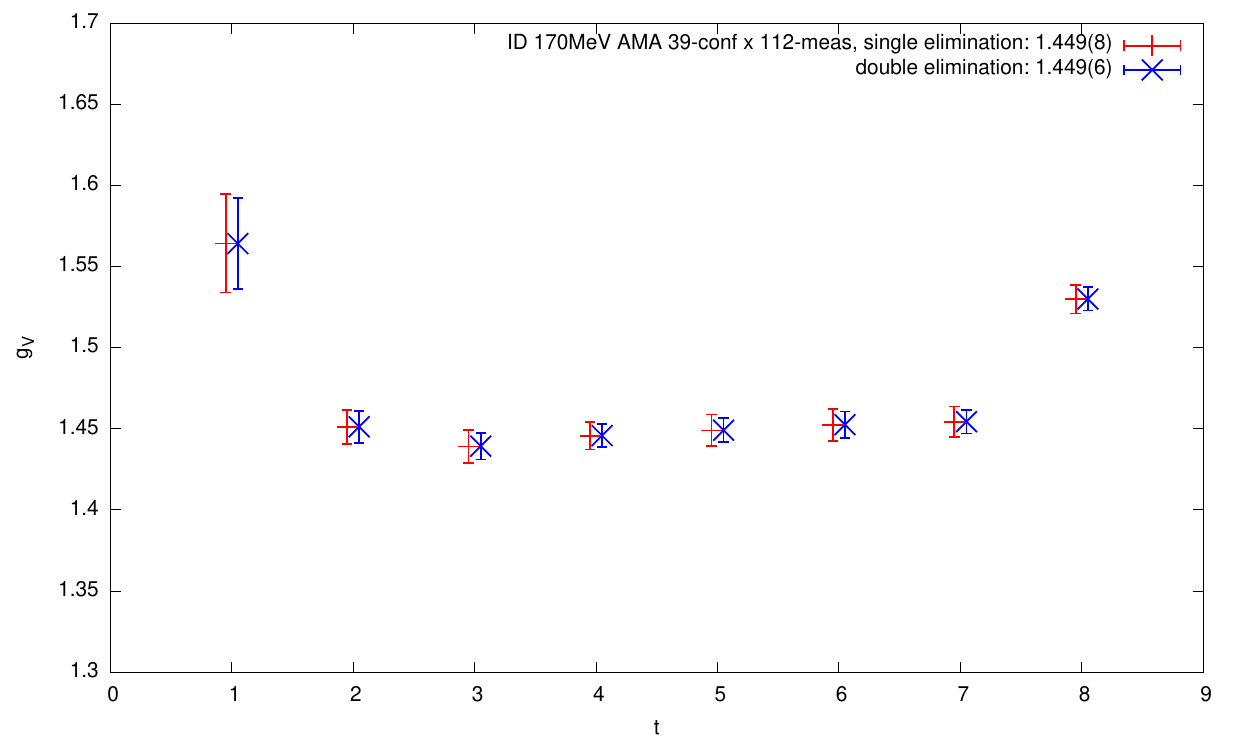}\\
\includegraphics[width=\columnwidth]{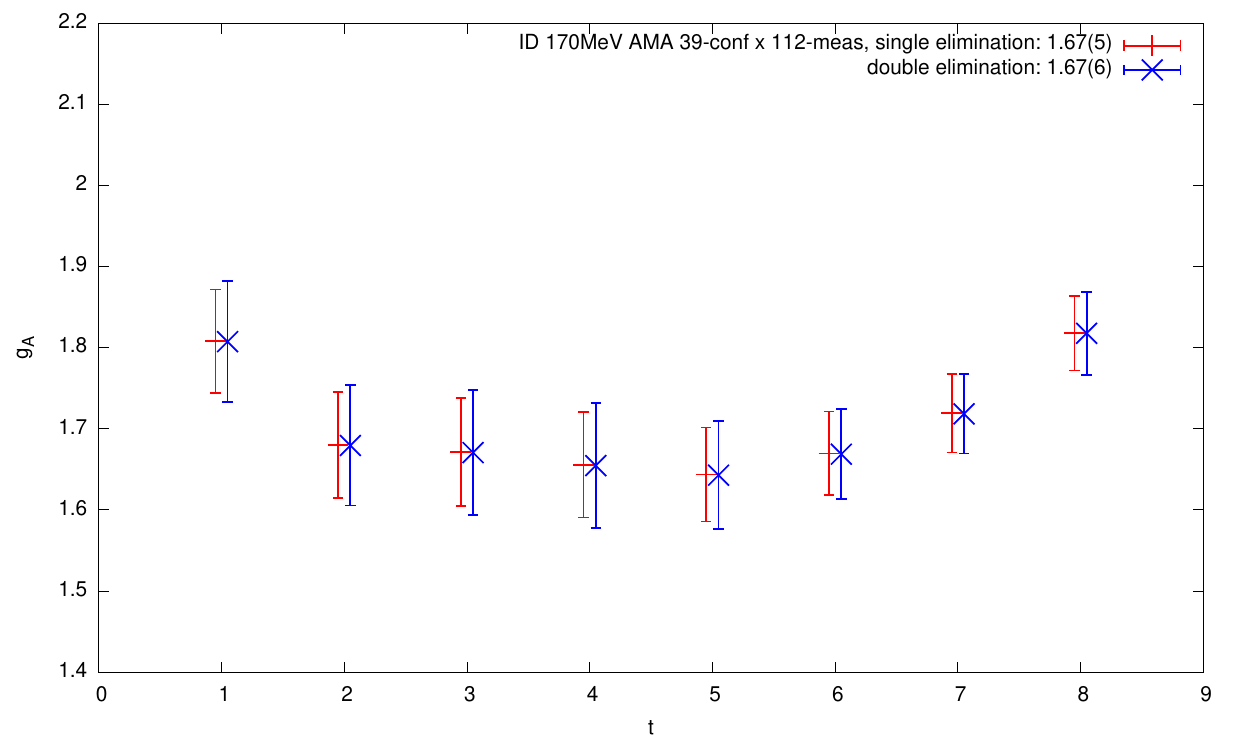}\\
\includegraphics[width=\columnwidth]{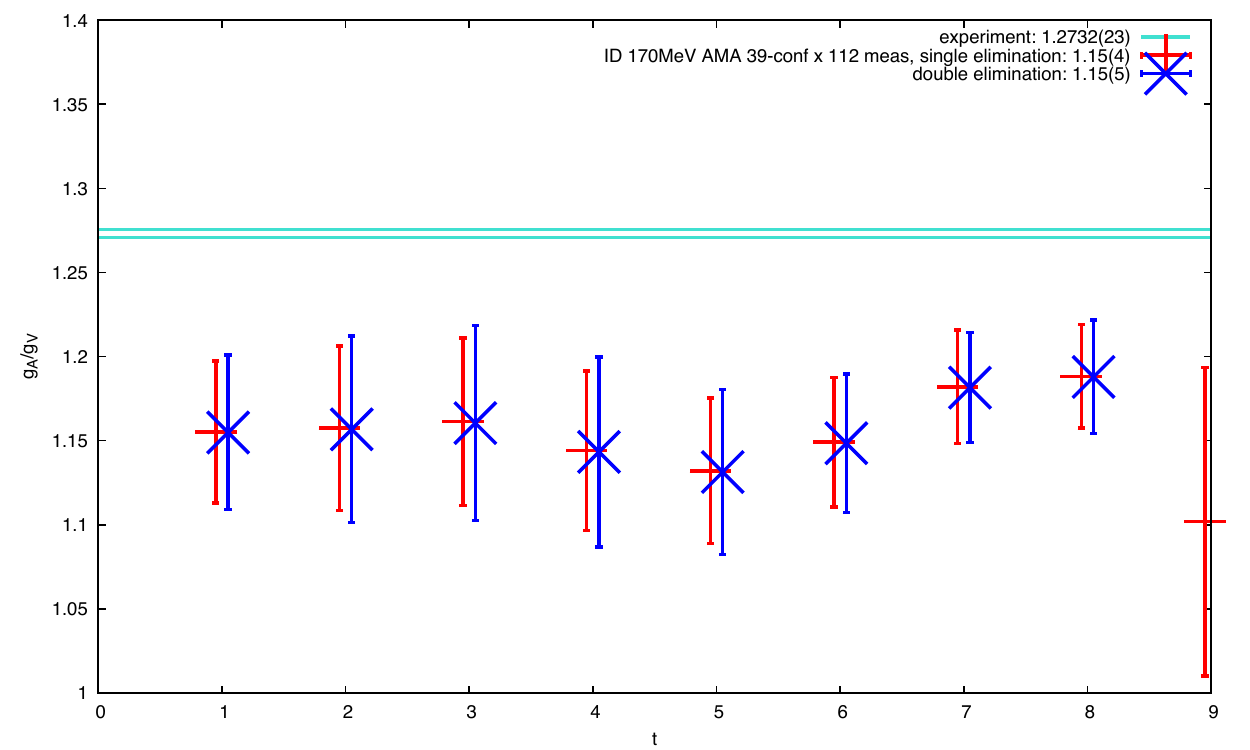}
\end{center}
\caption{
\label{fig:AVprecisions}
Comparative signal qualities of the isovector vector, \(g_V\), and axial, \(g_A\), charges and their ratio, \(g_A/g_V\), from the 172-MeV ensemble.
The charges are bare values from the local currents, and the ratio is naturally renormalized as the two currents share the common renormalization.
Clearly the ratio is more precisely determined than the axial-vector charge.
}
\end{figure}

As is the case for the vector current, we use the local-current definition for the axial-vector current.
Because the two local currents are connected by a chiral rotation, they share a common renormalization, \(Z_A=Z_V\), that relates them with the corresponding conserved global currents, up to \(O(a^2)\) discretization.
This is an advantage of the DWF scheme.
Thus for the axial-vector charge, \(g_A\), it is better to look at its ratio, \(g_A/g_V\), to the vector charge, for precision, as is demonstrated in Fig.\ \ref{fig:AVprecisions}.
The calculated value of the ratio, \(g_A/g_V\), underestimates the experimental value of 1.2732(23) \cite{PhysRevD.98.030001} by about 10\% and does not depend much on the pion mass, \(m_\pi\), in the range from about 418.8(1.2) MeV down to 172.3(3) MeV from four recent RBC+UKQCD 2+1-flavor dynamical DWF ensembles \cite{Allton:2008pn,Aoki:2010dy,Arthur:2012yc}.
The result appears to have been confirmed by several other major collaborations \cite{Dragos:2016rtx,Bhattacharya:2016zcn,Liang:2016fgy,Ishikawa:2018rew,Chang:2018uxx} using different actions but with similar lattice spacings and quark masses, though extrapolations to physical mass seem to differ.
It is especially important for calculations with Wilson-fermion quarks \cite{Dragos:2016rtx,Bhattacharya:2016zcn} to remove the \(O(a)\) systematic errors at the linear order in the lattice spacing, \(a\) \cite{Liang:2016fgy}. 
\begin{figure}[tb]
\includegraphics[width=\columnwidth]{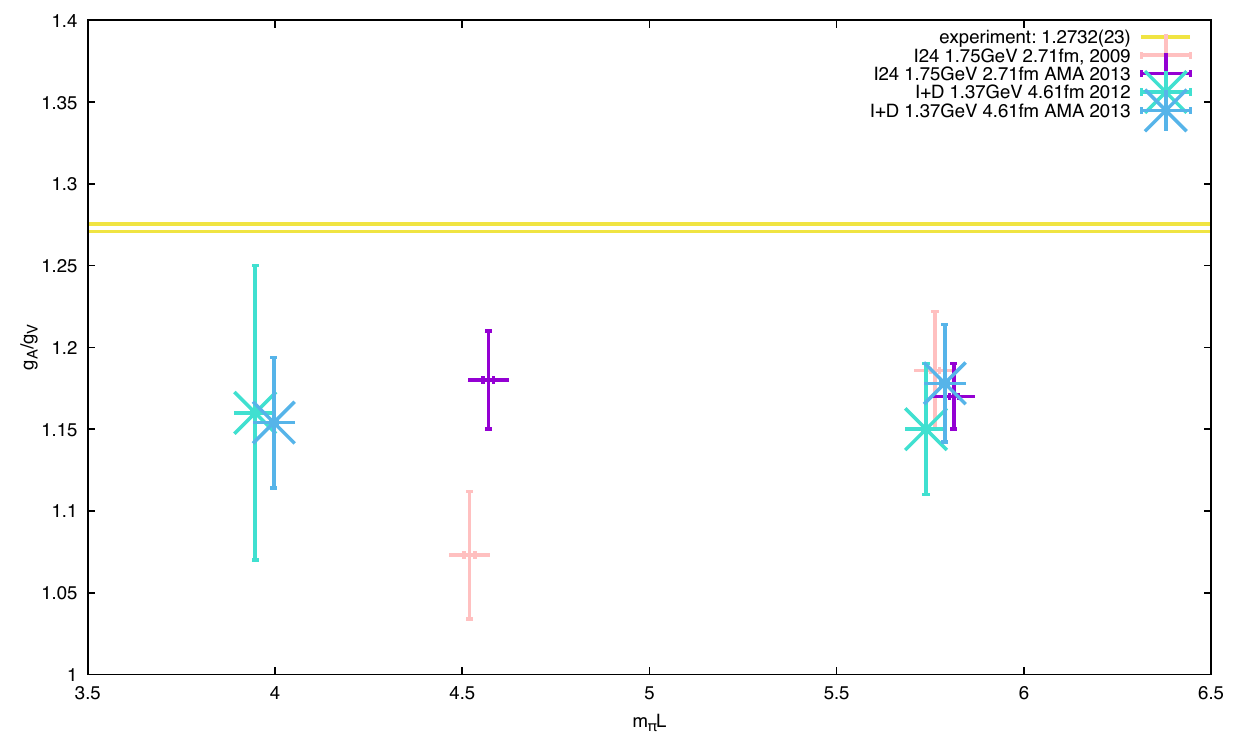}
\caption{\label{fig:AVmpiL}
Comparison of the dependence of the ratio, \(g_A/g_V\), of isovector axial-vector charge, \(g_A\), and vector charge, \(g_V\), calculated with recent RBC+UKQCD 2+1-flavor dynamical DWF ensembles, on the finite-size scaling parameter, \(m_\pi L\), in the present paper and our earlier reports.
The experimental value quoted here is 1.2732(23) from  the latest PDG \cite{PhysRevD.98.030001}.
}
\end{figure}

\begin{figure}[tb]
\includegraphics[width=\columnwidth]{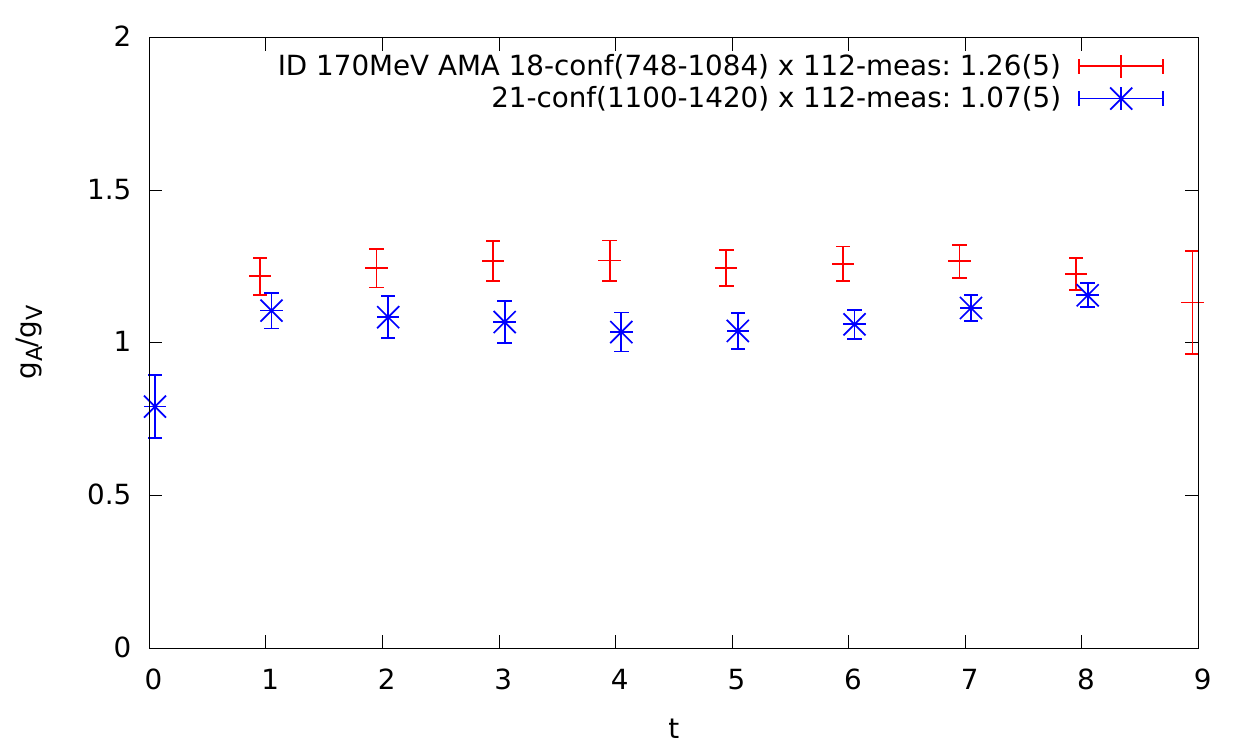}
\caption{\label{fig:gAgV2halves}
Plateaux of the ratio, \(g_A/g_V\), for the first (trajectory from 748 to 1084, red) and the second (1100 to 1420, blue) halves, respectively: fitted in the range from 2 to 7 lattice units, the values of 1.26(5) for the first and 1.07(5) for the second are almost four standard deviations away from each other.
}
\end{figure}

\begin{figure}[tb]
\includegraphics[width=\columnwidth]{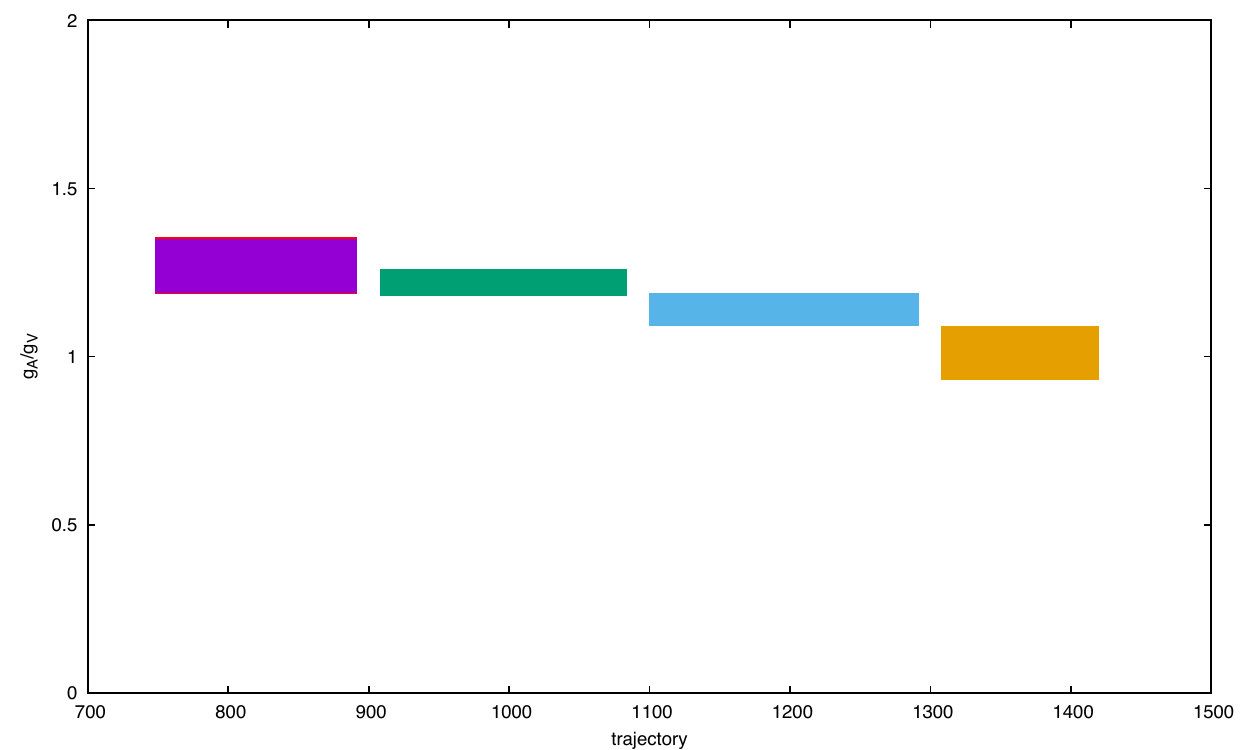}
\caption{\label{fig:gAgV4quarters}
Quarter-wise average along the hybrid MD time, from 748 to 892, 908 to 1084, 1100 to 1292, and 1308 to 1420: the values drift monotonically from what is consistent with experiment in the first quarter to a value around 1.0 in the last quarter.
}
\end{figure}

We are obviously suffering from some systematics that make our calculations undershoot the experimental value of \(g_A/g_V = 1.2732(23)\) \cite{PhysRevD.98.030001}.
Indeed we see possible signs of inefficient sampling:
First we observe an unusually long-range autocorrelation when we divide the lightest ensemble at \(m_\pi\) = 172.3(3) MeV into two halves, earlier and later, in hybrid MD time, as in  Fig.~\ref{fig:gAgV2halves}.
Indeed when we further divide into four consecutive quarters in MD time, the axial-vector charge starts at a value consistent with experiment but monotonically decreases to a value below unity, as in Fig.\ 
 \ref{fig:gAgV4quarters}.

Importantly, we also note that no such undersampling is seen in any other isovector observables we have looked at, including the vector charge, \(g_V\), quark momentum fraction, \(\langle x \rangle_{u-d}\) and quark helicity fraction, \(\langle x \rangle_{\Delta u-\Delta d}\), and that blocked-jackknife analyses with block size of 2 and 3 show strong correlation of two successive gauge configurations for \(g_A\) and \(g_A/g_V\).
Somewhat weaker autocorrelations may be seen in some other observables to block size of 2,  but then disappear at block size 3.

\begin{figure}[tb]
\includegraphics[width=\columnwidth]{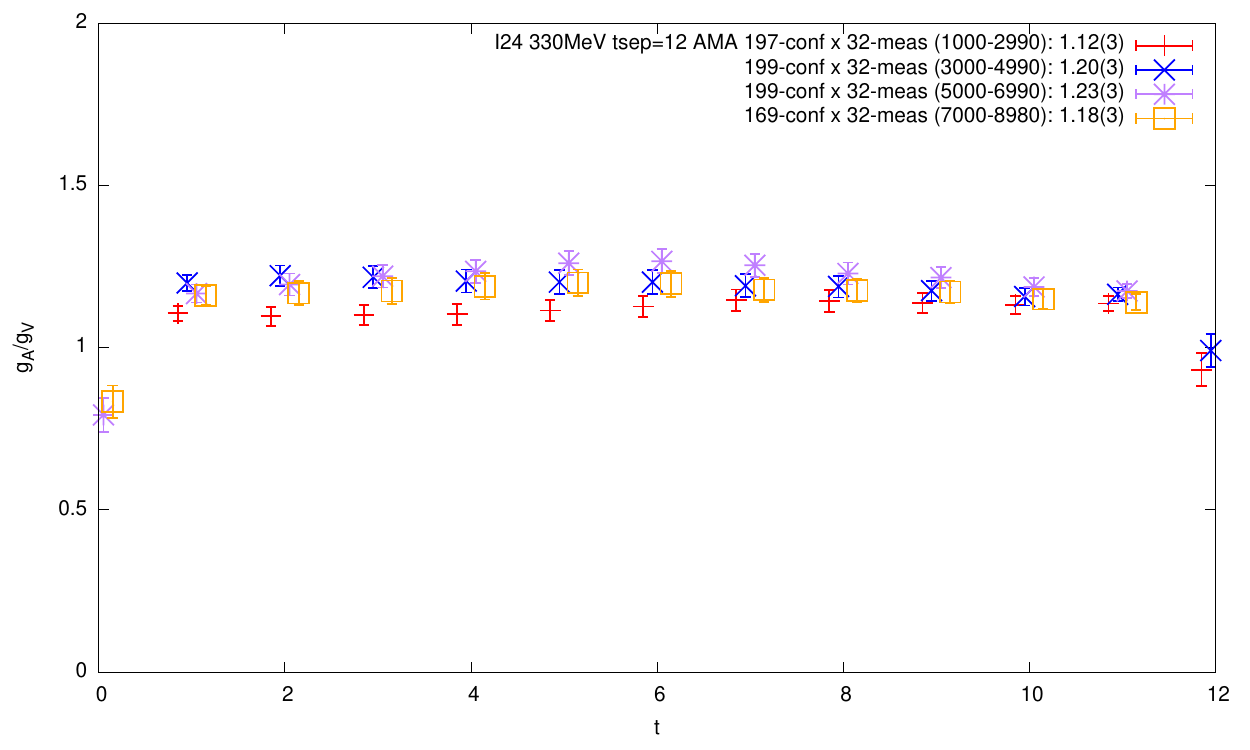}
\caption{
\label{fig:gA2gV330AC}
A weaker but similar long-range auto-correlation is also seen in \(g_A/g_V\) in \(m_\pi = 331.3(1.4)\) MeV ensemble when divided into quarters.}
\end{figure}

A similar but weaker sign of unusually long-range autocorrelation can be seen in the lightest of our earlier  ensembles \cite{Allton:2008pn} at \(m_\pi\) = 331.3(1.4) MeV when we divide it into four consecutive quarters in hybrid MD time as in Fig.\ \ref{fig:gA2gV330AC}.
However no such sign of undersampling is seen in the other two ensembles, the present one at \(m_\pi\) = 249.4(3) MeV and another at 418.8(1.2) MeV from the earlier work \cite{Allton:2008pn}.
In other words, the strongest sign of undersampling is seen at the smallest finite-size scaling parameter, \(m_\pi L \sim 4.00(6)\).
Another weaker indication of a long-range auto-correlation effect is seen at the second smallest but not the second lightest at \(m_\pi L \sim 4.569(15)\).
No effect is seen for larger values at \(m_\pi L \sim 5.79(6)\) or 5.813(12).
This of course does not prove that the problem is caused by the finite lattice spatial volume, but suggests so.

That there is a long-range autocorrelation in this observable is corroborated by blocked-jackknife analysis with block sizes of 2, 3, and 4, as is summarized in Tab.\ \ref{tab:blockJK}:
\begin{table}
\begin{center}
\begin{tabular}{cllll}
\hline
\multicolumn{5}{c}{Blocked jackknife analysis}\\
 & \multicolumn{4}{c}{block size}\\
 & \multicolumn{1}{c}{1}&\multicolumn{1}{c}{2}&\multicolumn{1}{c}{3}&\multicolumn{1}{c}{4}\\
\hline
\(g_V\) & 1.447(8) & 1.447(6) & \multicolumn{1}{c}{-} & \multicolumn{1}{c}{-} \\
\(g_A\) & 1.66(6) & 1.66(7) & 1.66(8) & 1.65(4) \\
\(g_A/g_V\) & 1.15(4) & 1.15(5) & 1.15(6) & 1.14(3) \\
\(\langle x \rangle_{u-d}\) & 0.146(7) & 0.146(8) & 0.146(8) & \multicolumn{1}{c}{-} \\
\(\langle x \rangle_{\Delta u - \Delta d}\) & 0.165(9) & 0.165(11) & 0.165(10) & \multicolumn{1}{c}{-} \\
\(\langle x \rangle_{u-d}/\langle x \rangle_{\Delta u-\Delta d}\) & 0.86(5) & 0.86(4) & \multicolumn{1}{c}{-} & \multicolumn{1}{c}{-} \\
\(\langle 1 \rangle_{\delta u - \delta d}\) & 1.42(4) & 1.42(6) & 1.42(6) & 1.41(3) \\
\hline
\end{tabular}
\end{center}
\caption{
\label{tab:blockJK}
Summary of blocked jackknife analysis for some isovector observables.
The error does not grow with block size except for the axial-vector charge and tensor coupling.
}
\end{table}
the statistical error of the axial-vector charge keeps growing to at least beyond a block size of 3 while those for the other observables do not grow at all except perhaps for tensor coupling which nonetheless stops growing earlier.

\begin{figure}
\begin{center}
\includegraphics[width=\columnwidth,clip]{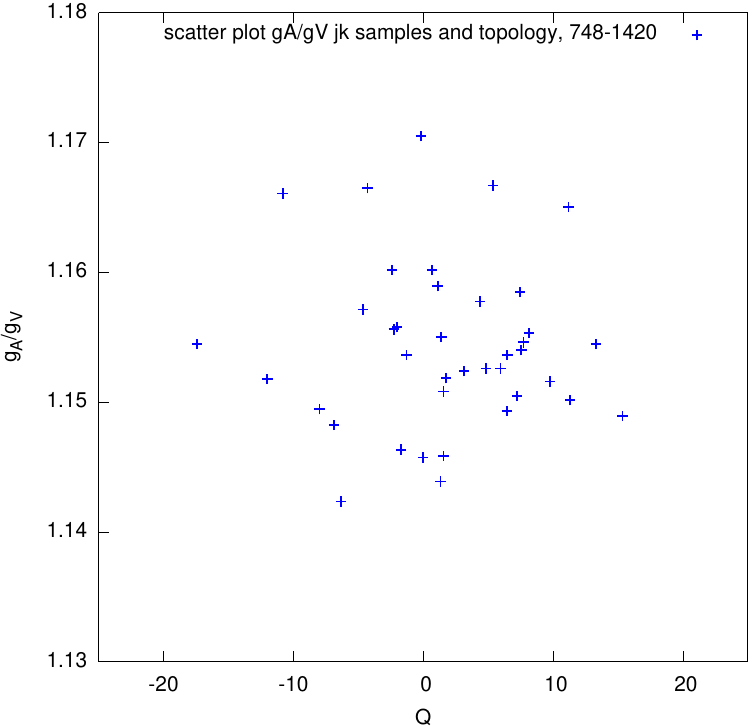}
\caption{\label{fig:gA2Topology}
Scatter plot of \(g_A/g_V\) jackknife samples against gauge topological charge:
No correlation is seen.
}
\end{center}
\end{figure}

If an observable appears to have long-range autocorrelation, it would be interesting to look at its correlation with the topology of the gauge configurations.
We explored this possibility by plotting jackknife samples against the topological charge (see Fig.~\ref{fig:gA2Topology}),
and did not find correlation.

\begin{figure}
\begin{center}
\includegraphics[width=\columnwidth,clip]{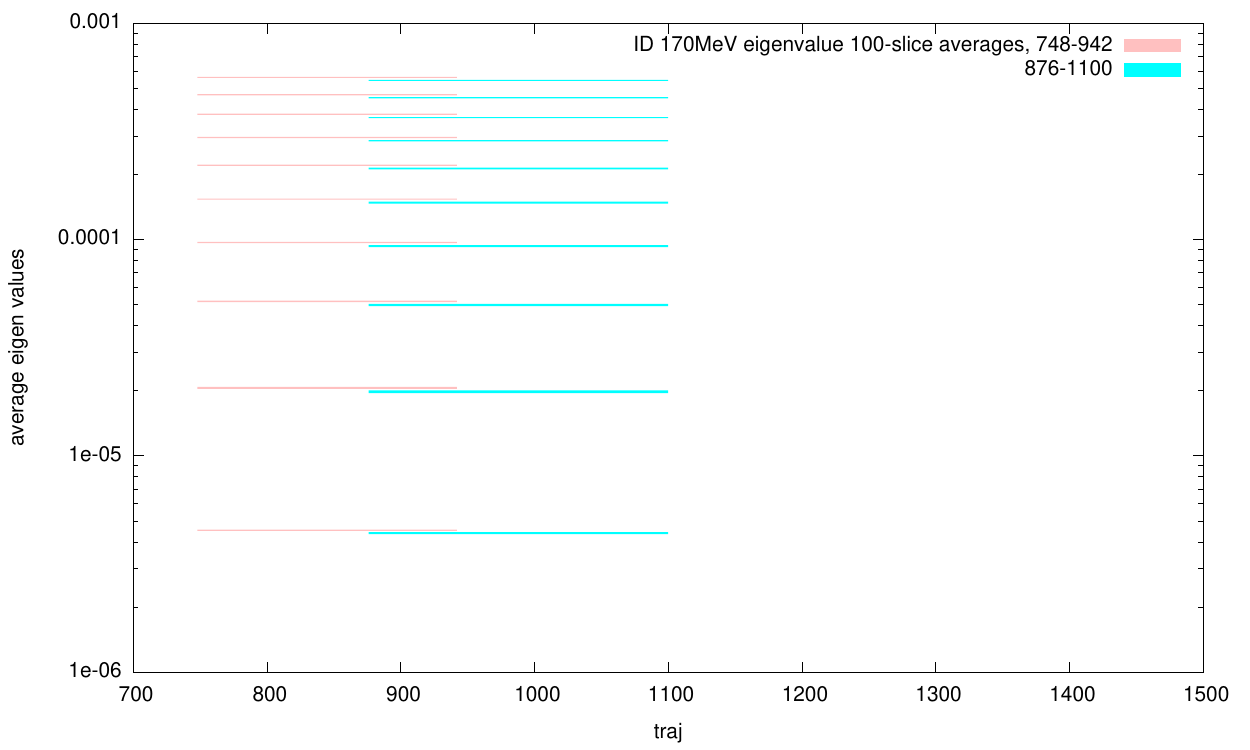}
\caption{\label{fig:gA2Deflation}
Deflation eigenmode statistics for the two quarters where data are available:
there is no difference in the lowest 100 modes averaged.
Some insignificant difference emerges in higher modes.
}
\end{center}
\end{figure}

We can also look at whether our low-mode deflation affected this, though the available information is limited to about half of the configurations of what we are presenting from the 172-MeV ensemble (see Fig.~\ref{fig:gA2Deflation}.)
Albeit with this limitation we do not find any correlation either:
average of the lowest 100 eigenmodes does not differ between the two halves.
Some difference emerges as we go to higher eigenmodes but does not appear to be significant.

On the other hand, as pointed out earlier in this paper, similar long-range autocorrelation was seen in the 331.3(1.4)-MeV ensemble \cite{Ohta:2013qda} that is at the second smallest \(m_\pi L\), but not in the lighter 249.4(3)-MeV nor the heaviest 418.8(1.2)-MeV ensembles with larger \(m_\pi L\), hinting that the systematics may arise from the finite-size effect.

It may be also instructive to remember earlier phenomenological analyses such as one done using the MIT bag model that estimates \(g_A/g_V = 1.09\) without pion \cite{Chodos:1974pn}, and another by the Skyrmion model that gives only a conditionally convergent result of 0.61 that is strongly dependent on pion geometry \cite{Adkins:1983ya}.

\begin{figure}
\begin{center}
\includegraphics[width=\columnwidth,clip]{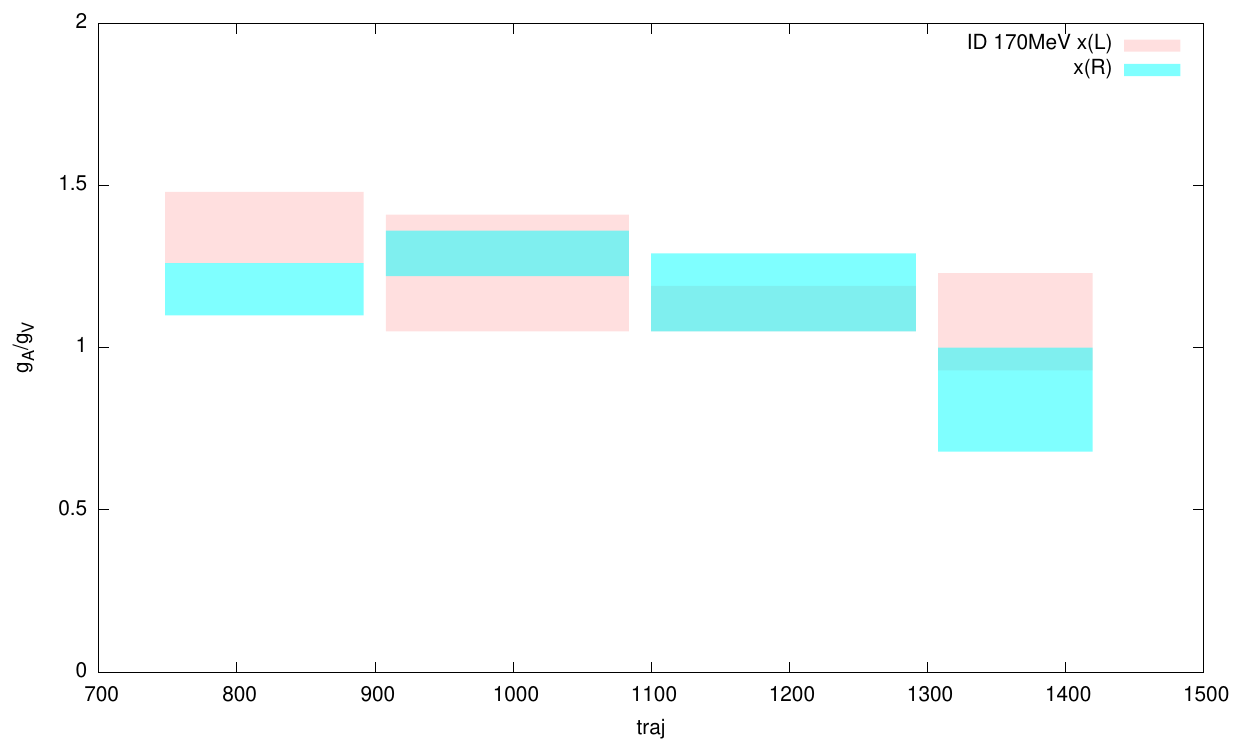}\\
\includegraphics[width=\columnwidth,clip]{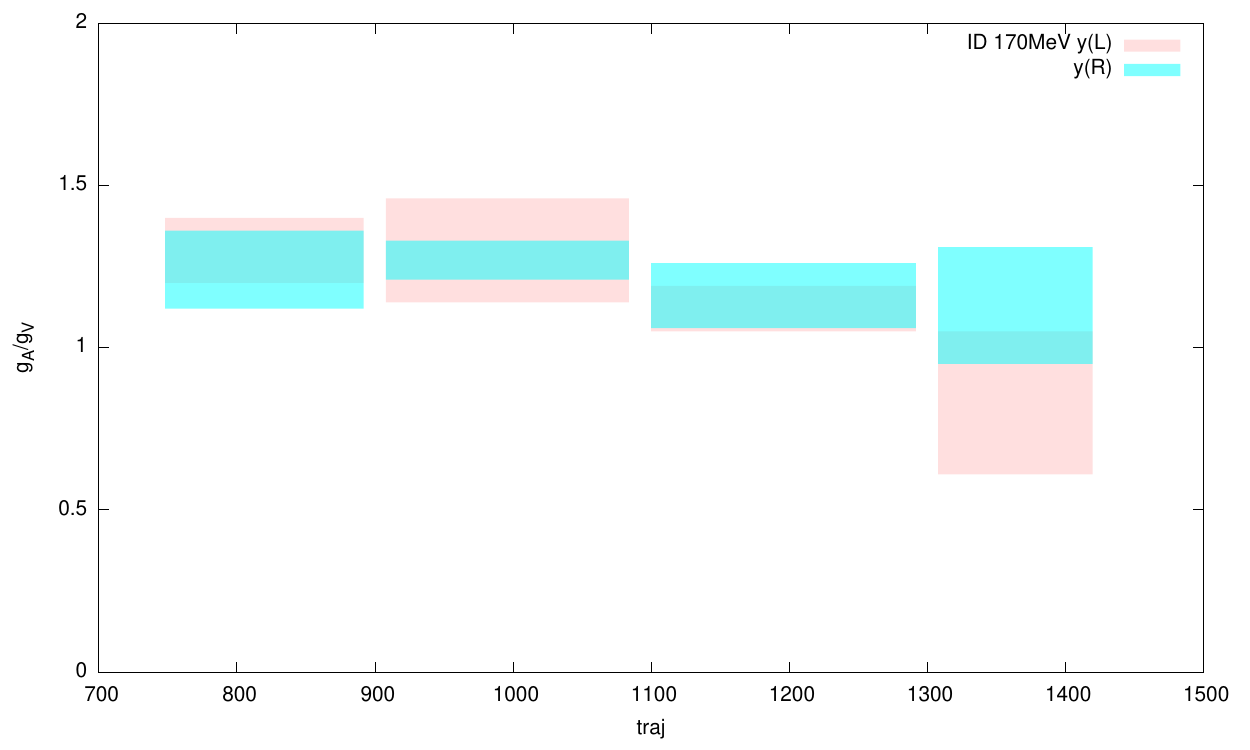}\\
\includegraphics[width=\columnwidth,clip]{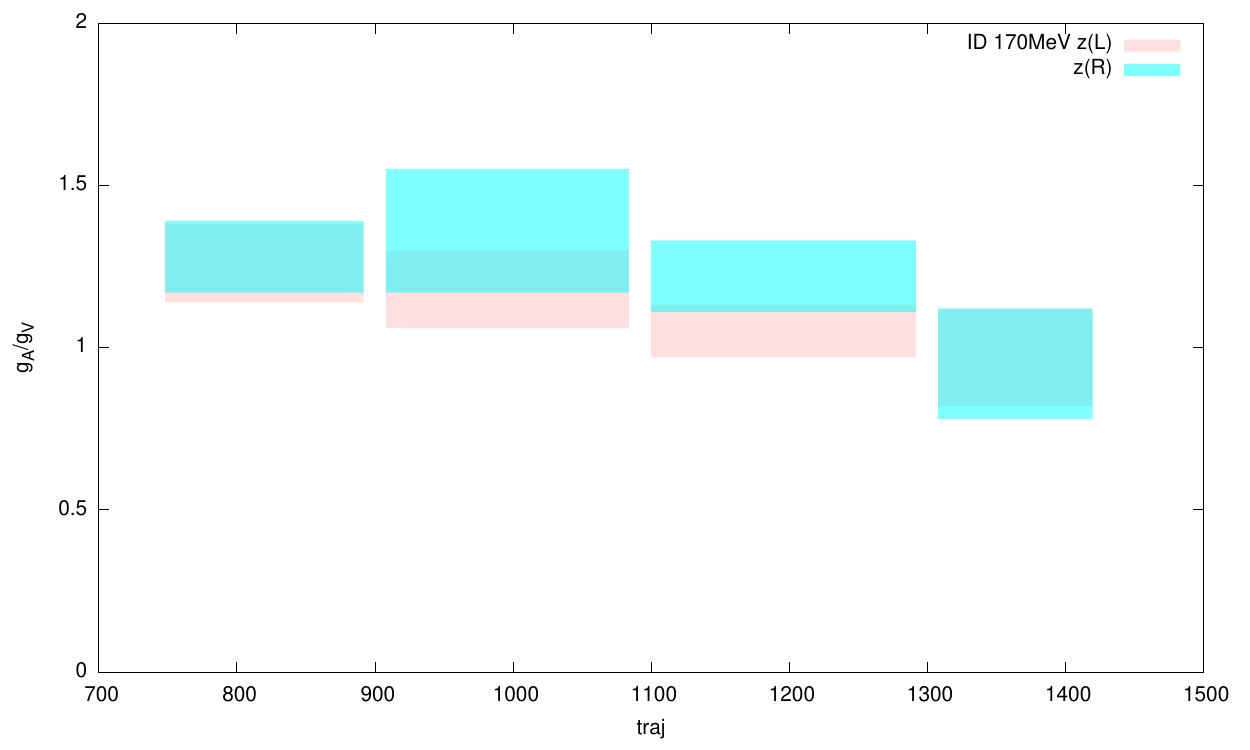}
\caption{\label{fig:piongeometry}
Evolution of the ratio, \(g_A/g_V\), along the course of molecular dynamics time, divided into two spatial halves in \(x-\) (top), \(y-\) (middle) and \(z-\) (bottom) directions.
The calculation appears to fluctuate spatially. 
Spin polarization is along the \(z-\)axis.
}
\end{center}
\end{figure}

To explore such spatial dependence arising from pion geometry, we divided the AMA samples into two spatial halves such as \(0\le x < L/2\) and \(L/2 \le x < L\) for each of the three spatial directions in order to check if there is any uneven spatial distribution (see Fig. \ref{fig:piongeometry}.)
We found that the calculation fluctuates in space.
Larger spatial volume would stabilize the calculation better.

\section{tensor and scalar couplings}

\begin{figure}[t]
\includegraphics[width=0.49\textwidth]{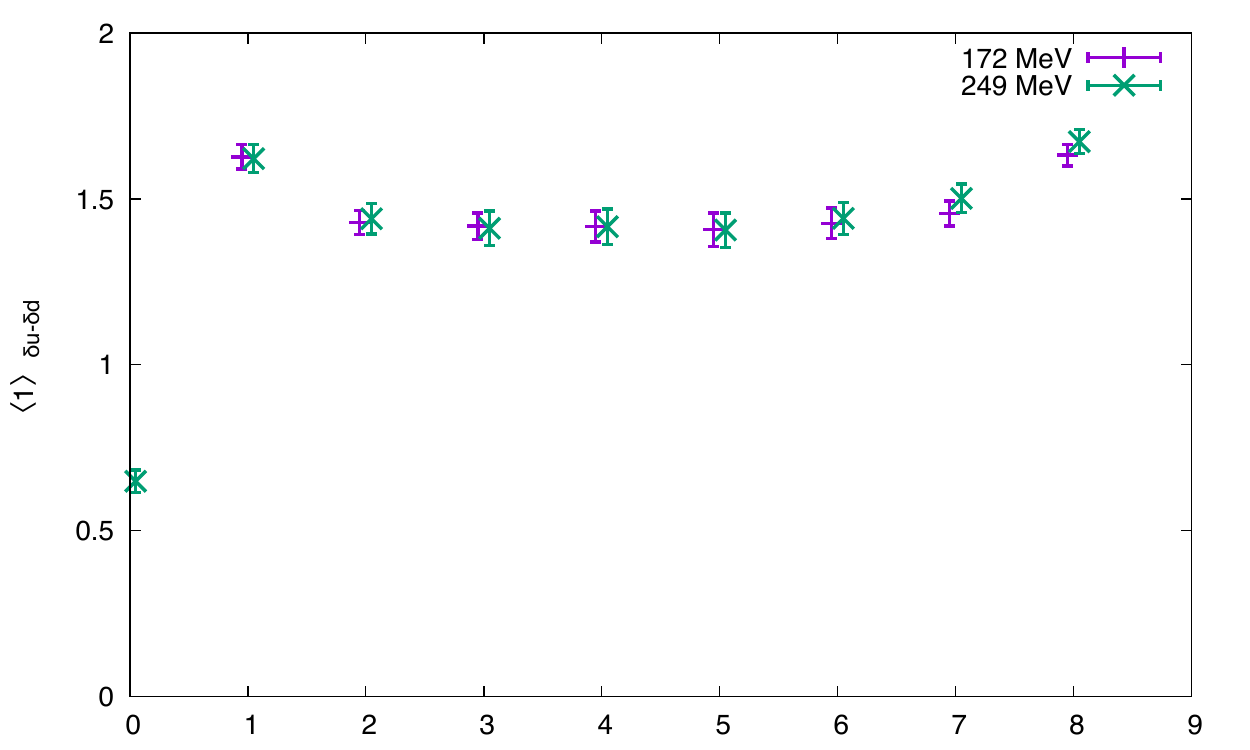}
\includegraphics[width=0.49\textwidth]{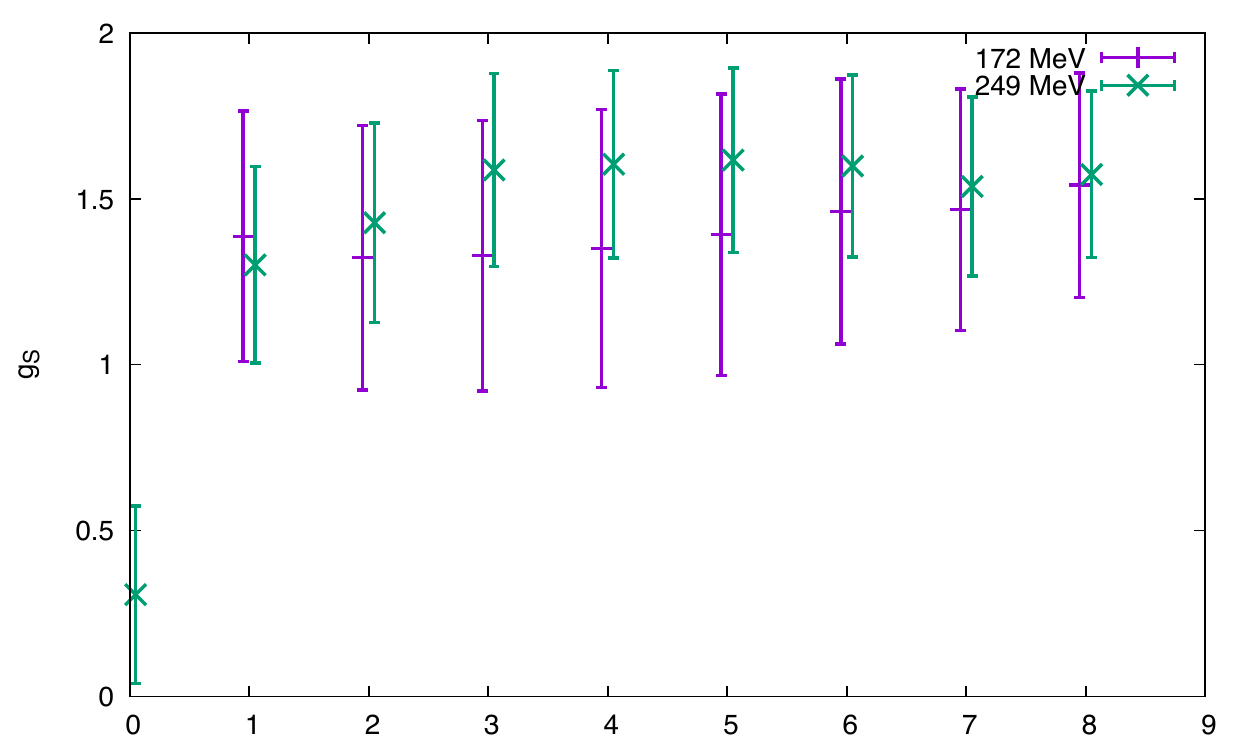}
\caption{\label{fig:1q}
Bare isovector tensor, \(g_T=\langle 1 \rangle_{\delta u - \delta d}\), and scalar, \(g_S\), coupling plateau signals.
The horizontal axes are lattice time in lattice units.
}
\end{figure}

Plateau signals for the bare isovector tensor, \(g_T=\langle 1 \rangle_{\delta u - \delta d}\), and the scalar coupling, \(g_S\), are presented in Fig.\ \ref{fig:1q}.
The tensor-coupling signals are very clean and do not show any mass dependence.
As was mentioned earlier, this observable shows weaker but still relevant signs of long-lasting autocorrelation similar to that of the axial-vector charge in the lighter, 172.3(3)-MeV, ensemble \cite{Ohta:2014rfa}.
Yet the agreement with the heavier ensemble where there is no such autocorrelation reassures this is less problematic in the tensor coupling than in the axial-vector charge.
The scalar plateaux are also well defined albeit with larger statistical errors.
No mass dependence can be seen here either.
In addition, the flatness of plateaux within respective statistical errors for both observables do not indicate measurable presence of excited-state contamination.

Bilinear operators are renormalized using the Rome-Southampton nonperturbative renormalization (NPR) method \cite{Martinelli:1994ty}.
This method allows determination of lattice matching factors (\(Z\)-factors)
to regularization independent (RI) schemes using lattice simulations directly, and without recourse to lattice perturbation theory.  
The validity of the method requires that the renormalization scale \(\mu\) is
sufficiently separated from the QCD scale and the scale of the lattice cutoff:
\begin{equation}  \label{npr-condition}
\Lambda_{\text{QCD}} \ll \mu \ll \frac{\pi}{a} \,.
\end{equation}
A perturbative calculation in the continuum is then conventionally used to
convert the RI-scheme results to the \(\overline{\text{MS}}\) scheme.

The relation between bare (lattice) operators and 
their renormalized counterparts is
\begin{equation}
\psi_{\text{R}} = Z_{q}^{1/2} \psi_{\text{B}}, \qquad \mathcal{O}_{\text{R}} = Z_{\mathcal{O}} \mathcal{O}_{\text{B}} \,.
\end{equation}
Here \(\mathcal{O}\) is one of \(S\), \(V\), or \(T\) for scalar,
vector, and tensor bilinears respectively.
We compute the amputated Landau-gauge fixed Green's functions of the operators 
of interest between external quark states with momentum \(p_1\) and \(p_2\).
Let these quantities be denoted \(\Lambda_{\Gamma}(p_1, p_2)\), 
where \(\Gamma\) is one of \(\{ S=1, V = \gamma^{\mu}, T= \sigma^{\mu \nu} \}\).
Here we use the RI/SMOM scheme defined in \cite{Sturm:2009fk},
for which the momenta satisfy \(p_1^2 = p_2^2 = (p_1 - p_2)^2\).
This differs from the original RI/MOM scheme proposal \cite{Martinelli:1994ty}
which instead takes \(p_1=p_2\).
This improved kinematic setup maintains a single renormalization scale,
but ensures momentum flows through the vertex thereby suppressing unwanted
infrared effects.
It was also found in \cite{Sturm:2009fk} that the RI/SMOM to \(\overline{\text{MS}}\) matching
factors have smaller \(\mathcal{O}(\alpha_s)\) coefficients.

There are two principal SMOM variants defined in \cite{Sturm:2009fk},
called RI/SMOM and \(\SMOMg\),
which differ in the definition of wavefunction renormalization.
The bilinear \(Z\)-factors are determined at the scale \(|p|\) 
by imposing the conditions
\begin{align}
\frac{1}{12} 
\Tr \left[ \Lambda_{S, \text{R}} (p_1, p_2) \right] &= 1  \notag 
\\
\frac{1}{144} 
\Tr \left[ \sigma_{\mu \nu} \Lambda^{\mu \nu}_{T, \text{R}}(p_1, p_2) \right] &= 1 
\label{renorm-cond} 
\\
\frac{1}{12q^{2}} 
\Tr \left[ q_\mu \Lambda^{\mu}_{V,\text{R}} (p_1, p_2) \, \Xsl{q} \right] &= 1 
 \qquad (\text{RI/SMOM}) \notag 
\\
\frac{1}{48} \Tr 
\left[ \gamma_\mu  \Lambda^{\mu}_{V,\text{R}} (p_1, p_2) \right] &= 1 
 \qquad (\SMOMg) \notag
\end{align}
on the amputated Green's functions in the chiral limit.
Here the traces are over both spin and color indices, 
and the normalization factors are such that
Eqs.~\eqref{renorm-cond} are satisfied by the bare operators in the free theory.

The propagators used to construct \(\Lambda_{\Gamma}(p_1, p_2)\) are computed using momentum sources, which results in very low statistical noise using a modest number of configurations.
Additionally, twisted boundary conditions on the quark fields allow us to continuously vary the magnitude of \(p\) while keeping the orientation fixed, resulting in smooth data as a function of renormalization scale. 

Results for the RI/SMOM and \(\SMOMg\) intermediate schemes are presented in Tabs.~\ref{tab:iresults}, \ref{tab:iresults2}, and \ref{tab:iresults3}.
\begin{table*}
\caption{
Scalar and tensor renormalization constants 
in the \(\text{RI/SMOM}_{(\gamma_\mu)}\) schemes at \(m_\pi = 172.3(3)\) MeV.
\label{tab:iresults}
}
\begin{tabular}{l  l l l l l l l }
\hline
\(\mu\) [GeV] & 1.53 & 1.63 & 1.72 & 1.82 & 1.91 & 2.00 & 2.11 \\
\hline\hline
\(Z_\text{S}/Z_\mathcal{V}\) (\(\SMOMg\)) & 0.813(1) & 0.829(2) & 0.842(2) & 0.855(1) & 0.866(1) & 0.878(1)& 0.888(1) \\
\(Z_\text{S}/Z_\mathcal{V}\) (RI/SMOM) & 0.892(1) & 0.906(2) & 0.918(2) & 0.930(1) & 0.940(1) & 0.951(1) & 0.960(1) \\
\(Z_\text{T}/Z_\mathcal{V}\) (\(\SMOMg\)) & 1.073(1) & 1.066(1) & 1.060(1) & 1.054(1) & 1.049(1) & 1.045(1) & 1.041(1) \\
\(Z_\text{T}/Z_\mathcal{V}\) (RI/SMOM) & 1.177(1) & 1.166(1) & 1.155(1) & 1.146(1) & 1.138(1) & 1.131(1) & 1.125(1) \\
\hline
\end{tabular}
\end{table*}
\begin{table*}
\caption{
Scalar and tensor renormalization constants
in the \(\text{RI/SMOM}_{(\gamma_\mu)}\) schemes at \(m_\pi = 249.4(3)\) MeV.
\label{tab:iresults2}
}
\begin{tabular}{l  l l l l l l l }
\hline
\(\mu\) [GeV] & 1.53 & 1.63 & 1.72 & 1.82 & 1.91 & 2.00 & 2.11 \\
\hline\hline
\(Z_\text{S}/Z_\mathcal{V}\) (\(\SMOMg\)) & 0.814(1) & 0.829(1) & 0.842(1) & 0.855(1) & 0.866(1) & 0.877(1) & 0.887(1) \\
\(Z_\text{S}/Z_\mathcal{V}\) (RI/SMOM) & 0.894(1) & 0.907(1) & 0.918(1) & 0.930(1) & 0.940(1) & 0.950(1) & 0.959(1)\\
\(Z_\text{T}/Z_\mathcal{V}\) (\(\SMOMg\)) & 1.073(1) & 1.066(1) & 1.060(1) & 1.054(1) & 1.049(1) & 1.045(1) & 1.041(1) \\
\(Z_\text{T}/Z_\mathcal{V}\) (RI/SMOM) & 1.178(1) & 1.166(1) & 1.156(1) & 1.146(1) & 1.138(1) & 1.132(1) & 1.126(1) \\
\hline
\end{tabular}
\end{table*}
\begin{table*}
\caption{
Scalar and tensor renormalization constants in the \(\text{RI/SMOM}_{(\gamma_\mu)}\) schemes in the chiral limit.
\label{tab:iresults3}
}
\begin{tabular}{l  l l l l l l l }
\hline
\(\mu\) [GeV] & 1.53 & 1.63 & 1.72 & 1.82 & 1.91 & 2.00 & 2.11 \\
\hline\hline
\(Z_\text{S}\) (\(\SMOMg\)) & 0.546(8) & 0.557(9) & 0.566(9) & 0.575(8) & 0.583(8) & 0.591(8) & 0.598(8) \\
\(Z_\text{S}\) (RI/SMOM) & 0.599(8) & 0.610(9) & 0.617(9) & 0.626(8) & 0.632(8) & 0.640(8) & 0.646(8)\\
\(Z_\text{T}\) (\(\SMOMg\)) & 0.722(8) & 0.717(8) & 0.713(8) & 0.709(8) & 0.706(8) & 0.703(8) & 0.700(8) \\
\(Z_\text{T}\) (RI/SMOM) & 0.792(8) & 0.785(8) & 0.777(8) & 0.771(8) & 0.766(8) & 0.761(8) & 0.757(8) \\
\hline
\end{tabular}
\end{table*}  
The wavefunction renormalization factors have been removed using \(Z_\mathcal{V}/Z_q\) determined from~\eqref{renorm-cond}, where \(Z_\mathcal{V}\) relates the local 4-d current to the conserved 5-d current.
These results are then converted to the \(\MSbar\) scheme using the two-loop perturbative expressions calculated in \cite{Gorbahn:2010kx,Almeida:2010ns}.

Taking the average of results from the two intermediate schemes after conversion to modified MS, and taking the full difference as an estimate of the systematic, we find for the renormalization factors of the scalar and tensor couplings:
\(
Z_S(\overline{\rm MS}, {\rm 2 GeV}) = 0.642(8)(22)
\)
and
\(
Z_T(\overline{\rm MS}, {\rm 2 GeV}) = 0.731(8)(24)
\)

From these, we obtain our estimates for the renormalized isovector tensor and scalar couplings as presented in Table \ref{tab:TS}.
\begin{table}
\begin{center}
\begin{tabular}{lll}
\hline
\multicolumn{1}{c}{\(m_\pi\) [MeV]} &
\multicolumn{1}{c}{\(g_T\)} &
\multicolumn{1}{c}{\(g_S\)}\\
\hline\hline
172.3(3)& 1.04(5) & 0.9(3) \\
249.4(3)& 1.04(5) & 1.0(2) \\
\hline
\end{tabular}
\end{center}
\caption{\label{tab:TS}
Renormalized isovector tensor, \(g_T=\langle 1 \rangle_{\delta u - \delta d}\), and scalar, \(g_S\),  couplings.
}
\end{table}
Neither is dependent on mass.

The tensor coupling is in good agreement with a value, about 1.0, obtained at the lightest mass in our previous calculations \cite{Aoki:2010xg} and also with later calculations by others \cite{Gupta:2018qil,Harris:2019bih}.
Its errors are dominated by a scheme-dependent systematics in non-perturbative renormalization, at about five percent, due mainly from the relatively low lattice cut off.

The scalar coupling, though noisier, is in broad agreement with other later calculations \cite{Gupta:2018qil,Harris:2019bih}.
The scalar errors are still dominated by statistical noise, but will eventually encounter the same scheme-dependent systematics in non-perturbative renormalization.

\section{Conclusions}

The nucleon mass calculated in the two present ensembles extrapolate linearly in pion mass squared, \(m_\pi^2\), to a value \(m_N = 0.950(5)\) GeV at the physical point.
This is to be compared with the average of proton and neutron mass experimental values, 0.938918747(6) GeV\cite{PhysRevD.98.030001}.
The slope in this linear extrapolation is steeper than that observed in our earlier calculations with heavier mass:
the result constrains non-linear dependence of nucleon mass on pion mass squared.

The isovector vector charge renormalizes to unity in the chiral limit.
This narrowly constrains excited-state contamination in the Gaussian smearing.

The ratio of the isovector axial-vector to vector charge shows a deficit of about ten percent.
This is in agreement with some other major lattice numerical calculations \cite{Dragos:2016rtx,Bhattacharya:2016zcn,Liang:2016fgy,Ishikawa:2018rew,Chang:2018uxx} using different actions but with similar lattice spacings and quark masses. 
The origin of this deficit is still to be understood.

We obtained good signals for isovector tensor coupling.
It does not depend on mass and extrapolates to 1.04(5) at physical mass with \(\overline {\rm MS}\) 2-GeV renormalization.
This is in agreement with the value obtained for the lightest pion mass of 331.3(1.4) MeV in our earlier work \cite{Aoki:2010xg} and also with calculations by others \cite{Gupta:2018qil,Harris:2019bih}.

Isovector scalar coupling is noisier but again does not show mass dependence, and is in agreement with other calculations \cite{Gupta:2018qil,Harris:2019bih}.

\section*{Acknowlegment}

The ensembles were generated using four QCDOC computers of Columbia University, Ediburgh University, RIKEN-BNL Research Center (RBRC) and  USQCD collaboration at Brookhaven National Laboratory, and a Bluegene/P computer of Argonne Leadership Class Facility (ALCF) of Argonne National Laboratory provided under the INCITE Program of US DOE.
Calculations of nucleon observables were done using RIKEN Integrated Cluster of Clusters (RICC) at RIKEN, Wako, and various Teragrid and XSEDE clusters of US NSF.
SO was partially supported by Japan Society for the Promotion of Sciences, Kakenhi grant 15K05064.

\bibliography{nucleon}

\end{document}